\documentclass[rnote]{aa} 
\usepackage{graphicx}
\usepackage{epstopdf}
\usepackage{txfonts}
\begin{document}
 \title{ 
Revised classification of the SBS carbon star candidates
including the discovery of a new emission line dwarf carbon star   
 \thanks {Based on observations made at the 1.52m telescope of the Bologna
 Observatory and  1.83m telescope of the Asiago Observatory.}  
  }
 \subtitle{  }
   \author{C. Rossi \inst{1}
   \and  K. S. Gigoyan \inst{2}
   \and M. G. Avtandilyan   \inst{3}
   \and  S. Sclavi. \inst{1} 
          }

   \institute{Department of Physics, University La Sapienza, Piazza A.Moro 00185, Roma, Italy\\
              \email{corinne.rossi@uniroma1.it}
         \and
  V. A. Ambartsumian Byurakan Astrophysical Observatory(BAO)and Isaac 
    Newton Institute of Chile, Armenian Branch, Byurakan 0213,
     Aragatzotn province, Armenia,\\
             \email{  kgigoyan@bao.sci.am }
             \and 
   Armenian State Pedagogical Uniiversity After Kh. Abovyan and Isaac  
      Newton Institute of Chile, Armenian Branch, Armenia. \\    
      \email{  mar\_avt@mail.ru }\\
             }

   \date{Received February 28, 2011 ; accepted }

 
  \abstract
   {  Faint high latitude carbon stars are rare objects commonly  thought to be  distant, luminous giants. For this reason they are often  used to probe the structure of the Galactic halo;   however  more accurate investigation of photometric and spectroscopic  surveys has revealed  an increasing percentage  of nearby objects  with  luminosities of main sequence stars.  }
   {   To clarify the nature of the ten carbon star candidates present in the General Catalog of the Second Byurakan Survey (SBS).   }
   {  We analyzed new optical spectra and photometry and used astronomical databases available on the web.   }
   { We verified that two stars are N$-$type giants already confirmed by other surveys.  We found that four candidates are M  type stars and confirmed the carbon nature of the remaining four stars; the characteristics of three of them are consistent with an early CH giant type. The fourth candidate, SBS\,1310+561 identified with a high proper motion star, is a rare type of dwarf carbon showing emission lines in its optical spectrum.  We  estimated absolute magnitudes and distances to  the dwarf carbon and the three CH  stars. {\rm ~}
   }
   { Our limited sample  confirmed the increasing evidence  that spectroscopy or colour alone are not conclusive  luminosity discriminants for CH$-$type  carbon stars .       }
   
   
   \maketitle
%

\section{Introduction}
  
The faint high latitude carbon stars (FHLCs , R $> $ 13, $|$b$| > $ 30$^{\circ}$) are rare objects. The observed spatial distribution indicates a surface density less than 0.06 deg$^{-2}$ (eg. Downes et al. \cite{down04} and references therein); for this reason they are seldom the specific target of dedicated photometric or spectroscopic surveys. 
It is of interest that  their colours are very similar to those of the much more numerous QSOs; in surveys aimed to search QSO, FHLCs are a small fraction of contaminants initially ''flagged'' as high redshift QSO candidates and then
identified as galactic objects by  their spectra  showing the characteristic absorption features of C2  Swan bands (eg. Totten et al. \cite{totten98},  Christlieb et al. \cite{christ01}, Downes et al. \cite{down04}). In this way  distant giants in the halo are usually discovered, but a non negligible fraction has a  significant proper motion indicative of a close object having the luminosity of a main sequence star. 

The discoveries of red faint stars from the prism surveys conducted at the Byurakan Astrophysical Observatory (BAO) are typical examples for this kind of follow up.
The results obtained by the first spectroscopic survey  conducted at BAO stimulated the need for a deeper survey aimed to reach fainter magnitudes  in searching for extragalactic objects with UV excesses and emission lines. The plates of 
the SBS were obtained at the BAO  between 1974 and 1991 with the 1m Schmidt telescope and three objective prisms in the range 7$^h$40$^m<\alpha<$\,17$^h$15$^m$ and 
 +49$^{\circ}<\delta<$\,+61$^{\circ}$  (Markarian \& Stepanian, \cite{mark83}, Stepanian et al. \cite{step90}); follow up spectroscopic programs with better resolution started soon after and, as a by-product a number of stellar objects were also found, mainly white dwarf (WD) and sdB subdwarfs. 
In some cases cool stars were also serendipitously discovered, the most
important being SBS\,1517+5017, showing a peculiar energy distribution. Higher resolution spectra revealed that the object is a rare type of binary system
composed by a quite hot white dwarf and a dwarf carbon star (Liebert et al. \cite{liebert94}).  The General Catalog of the SBS published by Stepanian
(\cite{step05}) is mainly dedicated to extragalactic objects but it also  includes  39  late type star candidates, listed in Table 41. Most of them simply classified  as red stars are probably M type stars; only ten were indicated as possible carbon stars. 

In the present paper we  clarify the nature of   these ten carbon  star  candidates. We  will discuss the characteristics of four new confirmed C-type stars with particular attention to SBS\,1310+561 which we discovered to be a dwarf carbon star (dC) showing emission lines in the spectrum, therefore being the most interesting object of this sample.


\section{SIMBAD association}

The  coordinates of the ten carbon star candidates published in the SBS General
Catalogue refer to the equinox 1950, with uncertainties  of the order of 1$-$2
arcmin,  making practically impossible the direct identification of the
objects  on the  Digital Sky Survey  maps. We precessed the coordinates
then, to check their carbon  rich nature, we started looking for these stars
in the Hamburg Quasar Survey  low resolution spectroscopic database (HQS, Hagen et al. \cite{hagen95})
in a field of 5 arcmin radius around the positions of each SBS object. From a closer inspection we verified that 
two of them, namely SBS\,0748+540 and SBS\,0832+534, were  also candidates from other surveys with different names and were already confirmed as N type giants on the basis of spectroscopic or photometric follow up (Knapik et al. \cite{knap99}). These two stars are already included in the General Catalogue of Galactic Carbon Stars (GCGCS, Alksnis et al. \cite{alks01}). 
Four of the remaining eight stars are in fact  M  type stars with  erroneous
classification as C stars in the SBS General Catalog. Two of them could be
identified with known M type variables, already included in the General
Catalogue of Variable Stars (X\,Uma is a Mira and AY\,Dra is a Semi Regular Variable star, Samus et al. \cite{samus10}). 
We   classified the other two  objects as M stars   with our  recent  observations described below.
The position of  SBS\,0854+530 corresponds to a variable star with no
previously published spectral classification (Maciejevski et
al. \cite{maci04}).  The last M  star, SBS\,1444+503 is  a faint object not associated with known sources. 

The remaining  four carbon star candidates  were not included in GCGCS; although their HQS spectra showed strong absorption bands, the resolution was not sufficient to derive more detailed information. We better clarified their  carbon nature and kinematical properties by the use of modern astronomical databases and by our recent dedicated observations described in the following sections.

In Table \ref{tab1} we give the revised data for all the ten SBS carbon candidates, including the four M and the two known carbon stars.

\begin{table}
\caption{Revised data for the ten C star candidates of the SBS General Catalog. }             
\label{tab1}      
\centering          
\begin{tabular}{l c c l l l }     
\hline\hline       
SBS & \multicolumn{2}{c}{Coordinates (GSC2.3)} & sp. & SIMBAD~~~~~~~~~~~ ref\\ 
number & $\alpha$ & $\delta$ & class &  association   \\ 
\hline                    
  0748+540  & 07 52 04.0  &   +53  56 54 & N & CGCS 1889 ~~~~~~~ [1] \\  
  0759+533  & 08  03  12.4 & +53  11  34 &  CH &    \\   
  0832+534  &08  36  27.6  & +53  17  52   &  N  & CGCS 2244~~~~~~~~~[1]  \\   
  0837+503  & 08  40  49.5  &    +50  08  12   &  M  &  X UMa ~~~~~~~~~~~~~~~[2]  \\   
   0854+530 & 08  57  44.3   &  +52  47  28  &  M9 &GSC03805-0109~[3]   \\   
1310+561   &  13  12  42.6   &  +55  55  54    & dC   &    \\   
  1444+503 &  14  46  36.4  &  +50  11 28    &   M4V &   \\   
  1537+571  &  15  38  39.3  &  +57  01  33    &  M  & AY Dra ~~~~~~~~~~~~~~[2] \\     
   1543+555 &  15  45  22.4  &   +55  21  33     &  CH  &   \\   
  1701+555  &  17  02  05.6   &  +55 27 13    &  CH &    \\   
\hline                  
\end{tabular}
\tablefoot{ Description of the columns:  1) SBS number according to Stepanian (\cite{step05}); 2) coordinates in the Guide Star Catalogue (Lasker et al.\cite{lask08})
   of the identified objects;  3) revised spectral classification;  4) other association according to SIMBAD database;
5) References:
[1] Alksnis et al. (2001); [2] Samus et al. (2010);  [3] Maciejevski et al. (2004).     }
\end{table}

\section { Recent observations}

We observed the four C star candidates  on 18 and 19 January 2010 with the
1.52\,m Cassini telescope of the Bologna  Astronomical Observatory equipped
with the Bologna Faint Object Spectrometer and Camera BFOSC, and a  EEV
P129915 CCD detector in the spectroscopic and photometric ($B,V, R$ Johnson
bands)  modes. We obtained moderate  resolution spectra for all the stars in
the range 3700$-$8500\,\AA \,(grism~\#4, dispersion 3.9\,\AA/pixel). We also obtained spectra in the range 4000$-$6600\,\AA \,(grism~\#7, dispersion 1.8\,\AA/pixel) for SBS\,0759+533 and SBS\,1310+561. We reduced all the data by means of standard IRAF procedures. 
  In Table \ref{tab2} we present  the averages of our photometry (3 values for each band); the corresponding ''rms'' from the individual values range between 0.05 and 0.09 mag,  quite good in spite of the fact that the nights were not excellent for photometric observations.
   In the last column we report the Galactic colour excess in the direction of the stars according to the extinction maps by Schlegel et al. (\cite{sch98}).
 We finally obtained  a spectrum  of our  M candidate SBS\,0854+530 with the
 same equipment  with grism~\#4 on 6 February 2011
and a spectrum of SBS\,1444+503  on 9 February 2011 with the 1.82\,m telescope of the Asiago Astronomical Observatory equipped with the Asiago Faint Object Spectrometer and Camera AFOSC and a TK1024 CCD and  grism~\#4,  with a dispersion 
4.8\,\AA/pixel in the range 3900$-$8500\,\AA.

\onlfig{1}{
  \begin{figure*}
   \centering
\includegraphics[width=\textwidth]{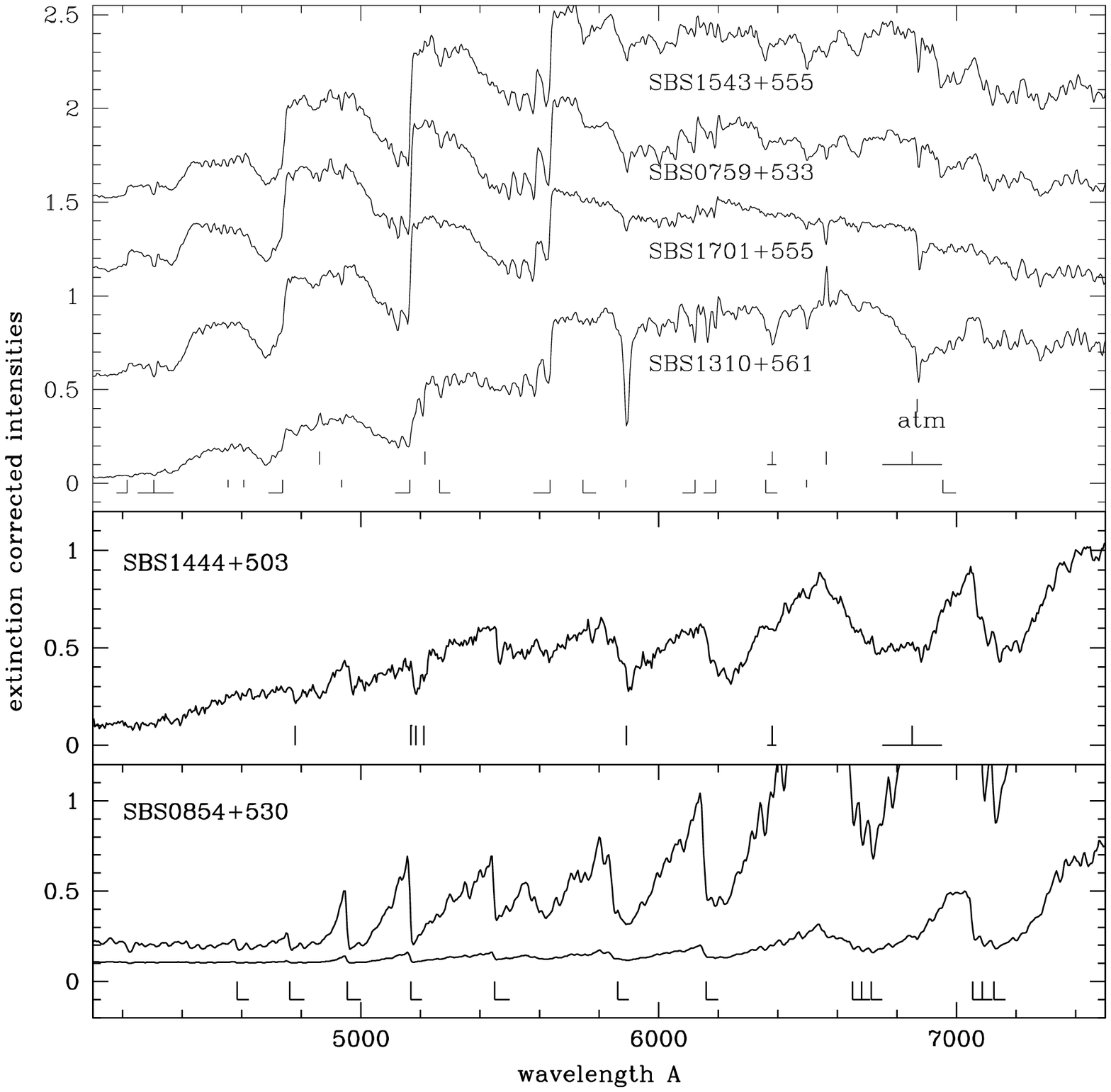}
 \caption{  CCD spectra of the recently classified SBS stars in the range 4100$-$7500\AA.  
 Bottom panels: the two confirmed M stars; bars in the lowest panel indicate TiO bands common to both stars  and VO bands of the late type;  lines prominent in SBS\,1444+503 are only reported in the corresponding panel.   Upper panel: the dwarf carbon and the three CH stars;
 vertical bars indicate  prominent features listed in the text. Bars corresponding to lines of particular interest for SBS\,1310+561 are vertically shifted of 0.15 units above the others.  }
  \label{figtot}
   \end{figure*}
  }  
%


\begin{table}
\caption{CCD optical photometry of our confirmed carbon stars.}             
\label{tab2}      
\centering          
\begin{tabular}{c c c c c c     }     
\hline\hline       
SBS & GSC2.3 & $B$ & $V$ & $R$  & E($B-V$)\\ 
number & number &   mag & mag & mag & mag        \\
\hline                    
  0759+533  & N8T4000463  &  14.15 & 13.23 & 12.50 &   0.034  \\   
  1310+561  & N4BP000199  &  16.6  & 14.76 & 14.00 &   0.015  \\   
  1543+555  & N4G7000252  &  16.42 & 14.83 & 14.20 &   0.011  \\   
  1701+555  & N4ID000249  &  16.24 & 15.19 & 14.12 &   0.016  \\   
\hline                  
\end{tabular}
\end{table}

\section{ Data analysis }

\subsection{  Spectral classification } 

The spectra of  SBS0854+530 and SBS1444+503 confirm their nature of M type 
stars.
 SBS\,0854+530 presents features typical for very late giant, very similar to those of the M7.5 giant RX Boo; the most prominent absorptions belong to 
the \ion{TiO}{}  bands at 4761, 4954, 5167, 5448, 5862, 6159, 6700, 7055, 7600\AA\, and the \ion{VO}{} bands of the red system  with several bandheads in the range 
7334$-$7472\AA,  and  7851$-$7973\AA, seen only in late type stars.~
SBS\,1444+503 can be classified as a M4 dwarf similar to GJ402 and LHS486; its most prominent features are:
the \ion{TiO}{} absorption bands at 4954, 5167, 5448,  6700, 7100, 7600\AA; the Mg $b$ triplet 5167,5173,5184\AA;
   very strong NaD doublet; the \ion{CaOH}{} diffuse bands centered at  5550 and  6230\AA.
the  \ion{MgH}{} bands at 4780, 5211\AA; the  \ion{CaH}{} at 6382, 6908, 6946\AA.  
 We will study in more details these stars in a forthcoming paper dedicated to the other 29 late type star candidates of the General Catalog of the SBS.
   
\begin{figure*} 
   \centering
 \includegraphics [ width= 17cm , height=14.0cm]   {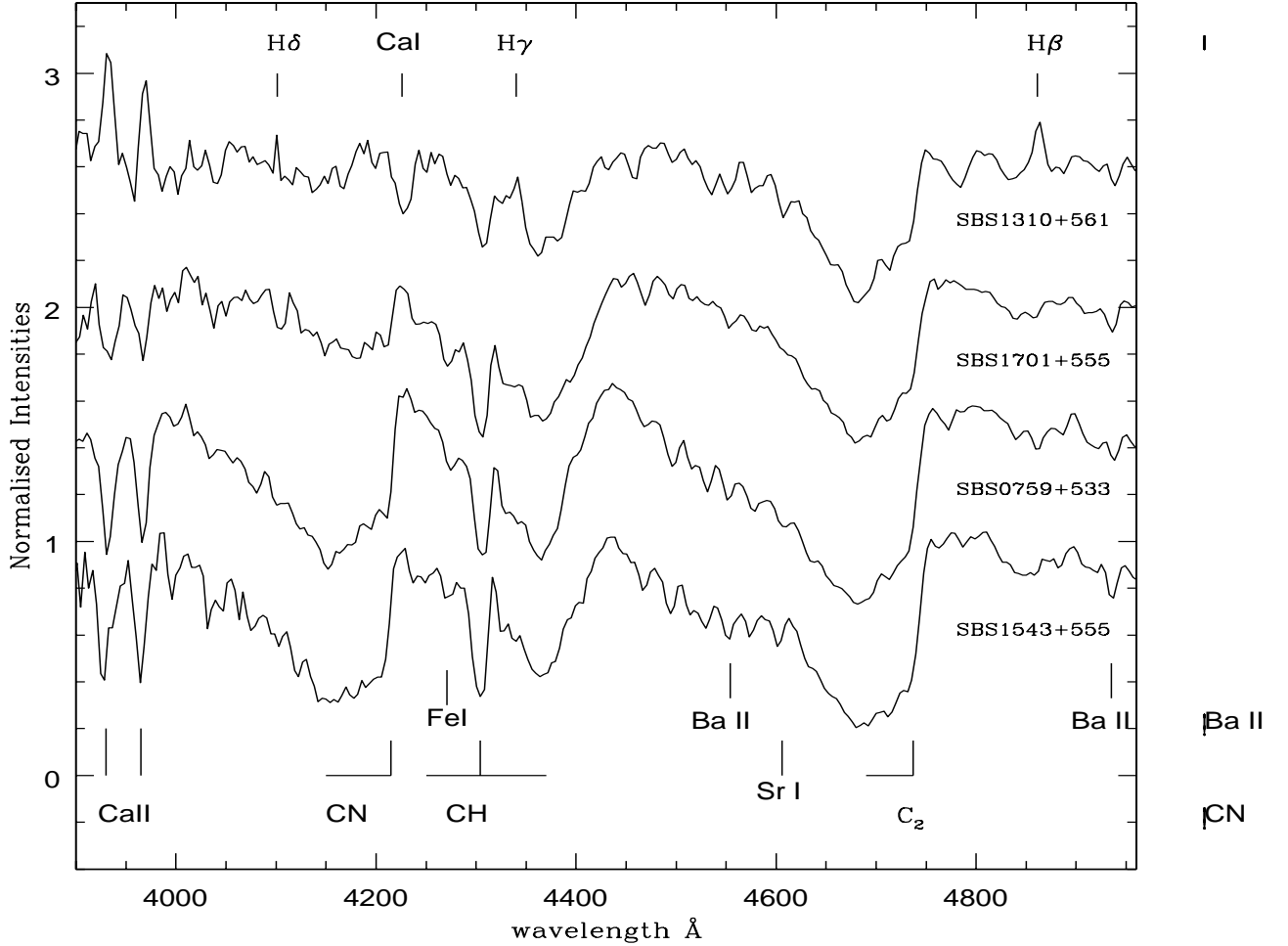}
 \caption{ Spectra of the four carbon stars in the wavelength region 3900$-$4960\,\AA, normalized to the continuum and vertically shifted. The most prominent features  are indicated.           }
         \label{figspenl}
   \end{figure*}

Concerning the four carbon candidates, the raw spectral classification from HQS was confirmed by our new spectra which allowed the determination of the carbon type class, finding in one case very interesting results. 
All the six   low resolution spectra in the range 3850-7600\,\AA,  are presented in Figure   \ref{figtot},  including those of the two M stars.
 The  spectra of the four carbon stars in the wavelength region 3900$-$4960\,\AA~ normalized to the continuum are shown in  Figure \ref{figspenl}. To be consistent with the data of
 SBS\,1543+555 and SBS\,1701+555 for which we only have grism~\#4 spectra,  and to show the region below 4000\,\AA ~
 the lower resolution spectra of SBS\,1310+561 and  of SBS\,0759+533  have been plotted in this figure.
 The better resolved  spectra of these two stars  are shown in  Figure \ref{figg7}.
We classified our stars on the basis of a number  of  spectroscopic and photometric characteristics. 
 At a first look the spectra could indicate either C$-$R or  CH type stars.
 Although our spectral resolution does not allow us to derive isotopic ratios,  in any case it is sufficient to detect  and measure the prominent  spectral features defining the carbon type  by comparing our targets with  those of known standards. 
 In  Figure \ref{figspenl}   important  lines and molecular bands   are indicated with vertical bars.       
Three stars, SBS\,0759+533, SBS\,1543+555 and SBS\,1701+555 are characterized by a number of  absorption features which can be summarized  as follows:\\
 Atomic lines :  \ion{Ca}{ii} K and H;   \ion{Fe}{i} 4271\,\AA \,;
 \ion{Ba}{ii} 4935, 4554, 6497\,\AA. The last  two absorptions  could have a contribution from close  \ion{CN}{}  lines, but the features  are  well visible also  in  the spectrum of  SBS\,1701+555 where the \ion{CN}{} bands  are faint and  the strength of the  isolated  4935\AA\,  line is similar in the three stars with  an equivalent width $\simeq$1.5 \AA ~  with respect to the local continuum.
 the \ion{Ca}{i} at 4226\,\AA \, is not detectable (marginally visible in the grism~\#7 spectrum of SBS\,0759+533).   \\
  Molecular bands:
 strong G band of \ion{CH}{}, with   prominent  secondary P branch at 4342\,\AA \,;  strong  $^{12}$C$^{12}$C bands at 4737, 5165, 5635, 6122, 6192\,\AA \,;  $^{12}$C$^{13}$C bands, if present are faint and not resolved. Anyhow   $^{12}$C$^{13}$C band head  at  4744\,\AA \, is marginally detected in the grism~\#7 spectrum of SBS\,0759+533.  The 
 \ion{CN}{}
  bands  at 4215, 5264, 5746,  6206, 6360\,\AA \,  are present in the spectra of SBS\,0759+533 and SBS\,1543+555. In SBS\,1701+555 only the band at 4215\,\AA \, is  strong, 
 other  \ion{CN}{}  bands   are barely visible or absent. 
 
 Considering the global appearance of the spectra one can see that
 with marginal differences in the  \ion{CN}{} bands SBS\,1701+555   matches quite well the spectrum of HD5223, used as CH standard by  Goswami et al. (\cite{gosw10}) and  classified as C$-$H\,3, C$_2$\,4.5, CH\,5 in Barnbaum et al. (\cite{barn96}). 
 The spectra of SBS\,0759+533 and SBS\,1543+555 are similar to that of  the CH star V\,Ari, classified  as C$-$H\,3.5, C$_2$\,5.5, CH\,4.5 in Barnbaum et al. (\cite{barn96}), but also to that of the late$-$R star RV Sct (Goswami et al. (\cite{gosw10}) and references therein). The  close inspection of the spectral lines and  the supplementary diagnostics of the infrared colors (see below)  have been fundamental in discarding this second hypothesis.\
  
 In summary,  the  spectroscopic and photometric characteristics of SBS\,0759+533, SBS\,1543+555 and SBS\,1701+555 led us to classify these stars  
 as  belonging to the CH class, with the differences  in the strength of CN bands  most probably depending on the chemical composition of the  atmospheres of individual stars (see   Wallerstein \& Kanpp \cite{wall98},  Lloyd Evans  \cite{llo2010}).  
 
%
The most interesting object is SBS\,1310+561 which  shows a number of differences with respect to the other stars: the Na\,D doublet is very strong (EW = 15.0$\pm$0.5\,\AA). \ion{Ca}{i} at $\lambda$ 4226\,\AA \,  and the molecular band at $\lambda$ 5211\,\AA\,
 of MgH are present with equivalent width of 3.5 and 4.1$\pm $0.2\,\AA, respectively. Noticeably all the Balmer lines are in emission; the \ion{Ca}{ii} doublet at $\lambda$ 3934, 3968\,\AA \, is also in emission
with EW= 7.0$\pm$0.4\,\AA ~and 6.0$\pm$0.5\,\AA. These values must be taken with caution given that owing to the low S/N ratio at the very blue end of the spectrum and   the spectral resolution we could not evaluate the contribution from the absorption components.
 While not common, emissions are not unexpected in carbon  giants (see for example Totten et al. \cite{totten00} and references therein), but in our case there are at least two indications for it  being  a main sequence object: one  is the high proper motion\, $\  \mu_{\alpha}=-118$ mas/y and  $\  \mu_{\delta}=+28$ mas/y (PPMXL catalog, Roesner et al.  \cite{roe10}); the other three stars have negligible proper motions in both directions. Another promising low luminosity good indicator for CH stars is the strength of the two CaH bands at $\lambda$\,6382 and $\lambda$\, 6750$-$6950\,\AA ~(Margon et al. \cite{margon02}). From a statistically significant sample of FHLCs those authors found these features strong only in the cool stars   having  detected proper motion. 
We measured an indicative value of the heliocentric radial velocity of the star by using the hydrogen emission lines of the higher resolution spectra; from Gaussian fits of H\,$_{\alpha}$  and H\,$_{\beta}$  we found -50  and -30 km/s with uncertainties of about 20 km/s.

\onlfig{3}  {
   \begin{figure*}
   \centering
   \includegraphics[width=\textwidth]{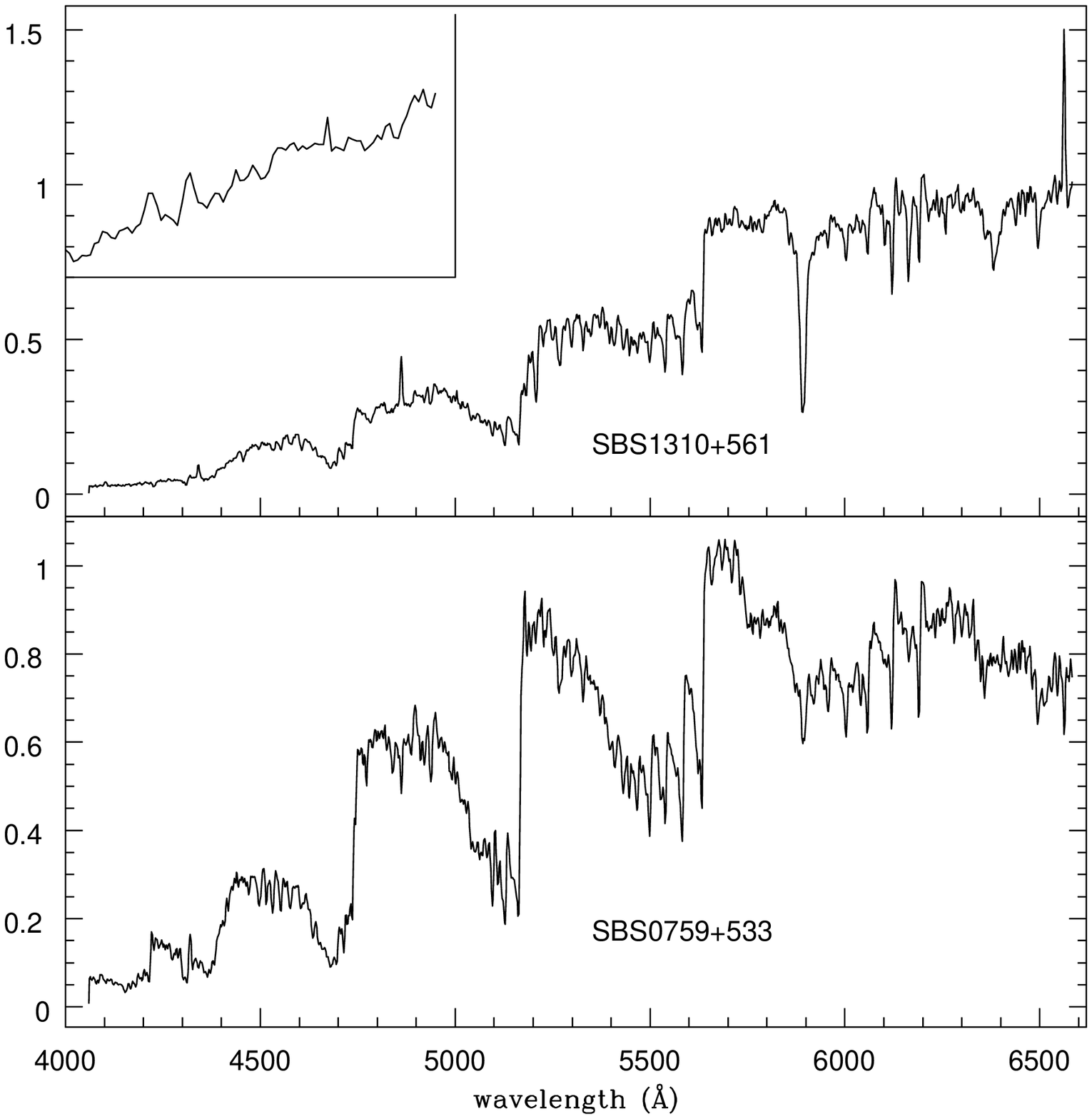}
      \caption{ Medium  resolution spectra of  the  CH giant SBS\,0759+533   and the dC SBS\,1310+561. For this star an enlargment of the region  3900 - 4200\,\AA \,  from the lower resolution spectrum  is  plotted to emphasize  H$\delta$    and the CaII emissions, 
 these last  being not covered by the grism~\#7   spectral range . Ordinates are the same as in Figure 1.
                    }
         \label{figg7}
   \end{figure*}
 } 
%

%
\begin{table*}
\caption{Infrared photometry and colours of the four  carbon stars. The last two columns are in the SAAO system. }             
\label{tab3}      
\centering          
\begin{tabular}{c c c c c c c c    }     
\hline\hline       
SBS & 2MASS & $J$ & $H$ & $K_s$ & $R-J$ & $J-H$ & $H-K$ \\ 
number &  identification & mag & mag & mag & mag & mag & mag    \\ 
\hline                    
  0759+533  & J08031240+5311340  & 10.84  &10.27   & 10.05 & 1.66 &0.66   & 0.16  \\
 1310+561   & J13124251+5555546   & 12.06  & 11.22 & 10.80 & 1.94 & 0.92  & 0.36  \\
   1543+555 & J15452240+5521327  &  12.57  & 11.85  & 11.62 & 1.63 &0.80  &0.19   \\
  1701+555  & J17020556+5527134  & 12.87 & 12.34   & 12.19   & 1.25 & 0.60  & 0.12  \\
\hline                  
\end{tabular}
\end{table*}

\subsection{Optical and infrared photometry  }

The optical colours are typical for CH-type stars with the dwarf being the reddest. We can assume that these colours are very close to the intrinsic values because at the Galactic latitudes of our stars the  interstellar reddening is negligible  (see Table \ref{tab2}).
All the data from our new photometry agree quite well with those of GSC2.3 within the errors, indicating that no major changes occurred between the different epochs of observations. For SBS\,0759+533, SBS\,1310+561 and SBS\,1543+555 there are also photometric data obtained during the Sloan Digital Sky Survey. The first star is saturated;
 for the other two stars the original photometry, converted to the Johnson bands with the transformation equations by  West et al. (\cite{west05}), yields the following magnitudes:
~for SBS\,1310+561:~$B=16.8, \  V=15.0, \  R=13.8$  and   for SBS\,1543,+555:
~ $ B=16.6,\   V=15.2, \   R=14.2$. Taking into account our measurement uncertainties and the fact that the SDSS conversions are not calibrated to carbon stars, the agreement is very good. 

\begin{figure}
  \centering
   \includegraphics[height=8cm  ]{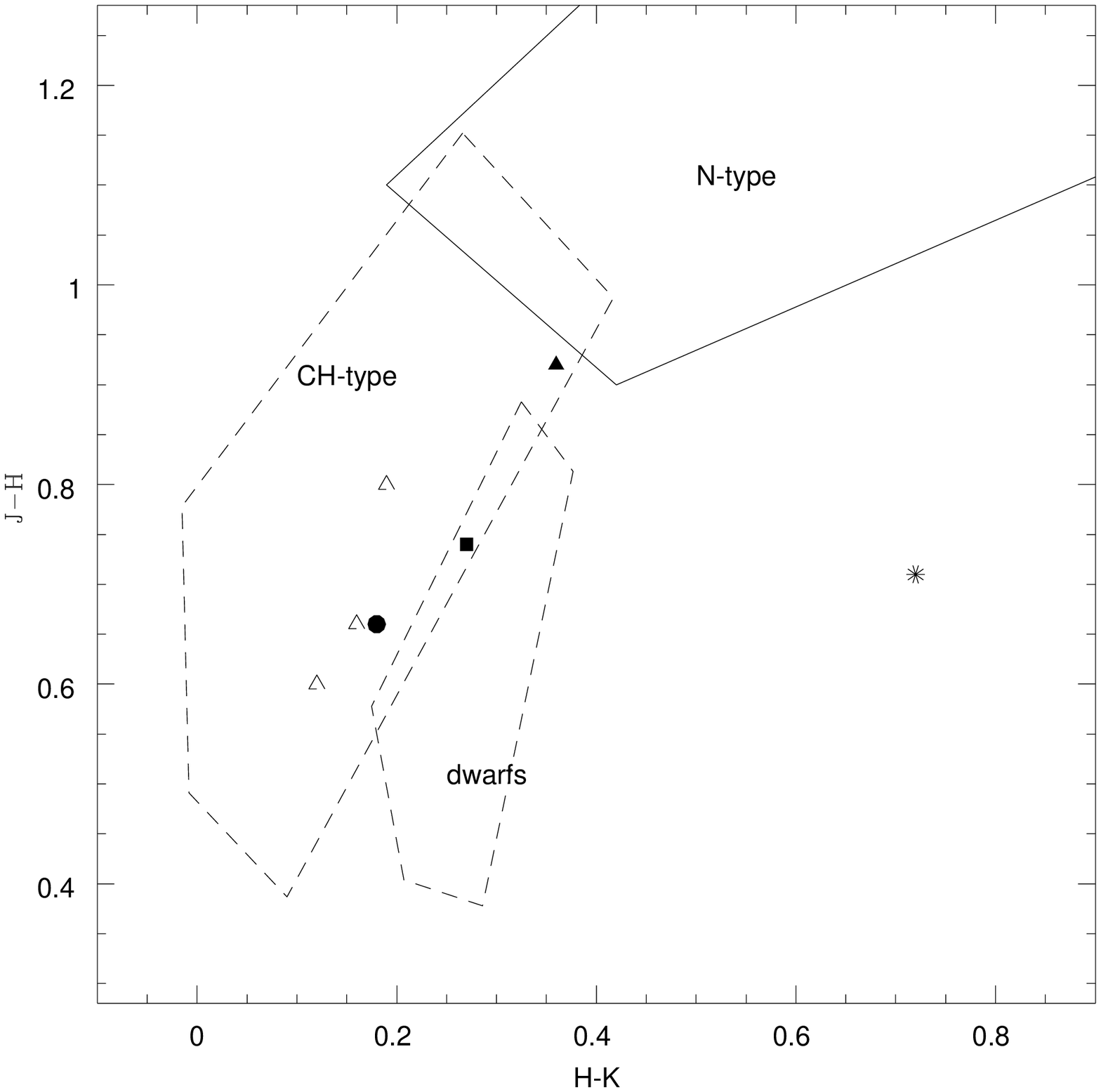}
   \caption{Near infrared two$-$colour diagram in the SAAO photometric system; the zones limiting the loci of the  different  carbon classes are 
those defined in  the Figure 3 by Totten et al. (\cite{totten00}).  Open triangles: the three SBS$-$CH giants; filled triangle: SBS1310+561; filled square: CLS29;  filled circle: PG0824+289;  star: SDSS\,J085853.3+012243.
              }
         \label{figcol}
   \end{figure}

From the infrared magnitudes published in the Two Micron All Sky Survey (2MASS) catalogue (Skrustkie et al. \cite{skru06}), we could check the position of our stars in the infrared colour$-$colour diagrams. In Table \ref{tab3} we present the 2MASS magnitudes and  $R-J$ ~colour. In the last two columns we report the colours $J-H$, $H-K$ in the SAAO photometric system obtained by transforming the  2MASS magnitudes according to the formulae by Koen et al. (\cite{koen07}). In the  infrared two$-$colour diagram   shown in  Figure\ref{figcol}   SBS\,0759+533, SBS\,1543+555 and SBS\,1701+555 are placed  inside the region occupied by the great majority of CH$-$type  stars studied by Totten et al  \cite{totten00}, colors confirming  the spectral classification. 
 For these stars the need for a spectroscopic classification  originates from the fact that  there is no obvious distinction in the  colors of CH and early  C$-$R stars.  We will  better discuss  this argument  in section 5. 

SBS\,1310+561 lies on the ill defined boundary between CH and dwarf zone. 
  Border line colours are not unique in the small family of the emission line dC. To our knowledge there are only three confirmed dC stars previously reported showing emission lines: CLS29, identified by Totten et al. (\cite{totten00}), which also lies on the boundary region between CH and dwarfs; PG0824+289 placed in the CH domain, has a composite spectrum (Heber et al. \cite{heber93}); in this case most probably the optically visible white dwarf companion affects the infrared emission of the system.  The opposite case is  the third star, SDSS\,J085853.3+012243 which is the reddest object of the sample studied by Downes et al. (\cite{down04}). 
The situation is  less clear for SBS\,0759+533, SBS\,1543+555 and SBS\,1701+555 by using directly the 2MASS colours and the more general diagrams of figures 2
 ($R-J$\, vs \,$J-K_s$) and 3 ($J-H$\, vs \,$H-K_s$) of Lowrance et al. (\cite{low03}).
But these authors remark that the tracks  for giants and dwarfs overplotted in their figure 3  were obtained by  interpolating the data of figure 5 of Bessel \& Brett (\cite{bessel88}), based on colours of {\bf non} carbon stars. Anyhow in these plots SBS\,1310+561 falls well inside the region defined by known dC stars.

The photometric behavior of all the four stars has  been monitored by the Northern  Sky Variability Survey (NSVS, Wozniak et al. \cite{woz04}) between March 1999 and May 2000.
 SBS\,0759+533 is constant with very little scatter from the mean magnitude.
 The other three stars show fluctuations of about $\pm$0.4 mag amplitude as shown in  Figure \ref{figlc}.
 It is worth to remember that the NSVS system has an overall response similar to the Red Johnson band but the values are measured from an unfiltered CCD detector, therefore the zero magnitude strongly depends on the stellar colours; for the considered stellar type the NSVS magnitudes are fainter of about 0.3-0.5 magnitudes with respect to the Johnson filter.

\section{ Discussion }
It is not necessary to repeat here the detailed discussion made by Margon et al. (\cite{margon02})  and Downes et al. (\cite{down04}) on the reliability of 
the spectroscopic and photometric luminosity discriminants for CH-type stars. {
It can be sufficient to remember  their final comment that at low dispersion 'spectra and colours of the disparate classes are frustratingly similar'. 
 The same consideration  is valid when  the various   types  of carbon stars are considered.  While infrared colors of N and late$-$R giants are well separated from CH and early$-$R stars ,  there is no clear distinction between  CH and early$-$R,  as we noted in section 4.2, nor between N and late$-$R types.
 This problem  has  been discussed in detail by Zamora et al \cite{zam09} who made an accurate  photometric and spectroscopic  analysis  of 17  previously  classified R stars.
 The high resolution spectra and the use of the most recent stellar models allowed these authors   to find that   40\% of  their sample  were wrongly classified; this result confirms previous doubts about the  real relative  number of R  and CH types among the carbon stars. 
The importance  from the evolutionary point of view  has been deeply discussed by Zamora et al \cite{zam09}; we refer to that paper for further reading.
 
We want to remark here that there are also a few minor, but very common causes of  confusion in the color plots:
one is the fact that  often data from  different photometric systems are put together without taking into account the necessary  transformations, increasing 
in this way the spread of the results; another one is that often previous schemes are strictly  adopted  without considering that the conditions are never tight and/or those schemes are based on possibly  different  photometric systems. In our case,  dealing also  with an emission line star we paid particular attention to  the photometric system used to place our stars in a two color diagram.  For the limited number of known emission line  dwarf  carbon the photometric situation is  further confused by the possible presence of a companion, as discussed below.  }

Also photometric fluctuations cannot be considered  conclusive luminosity indicators: CLS29 and PG0824+289 were also monitored by the NSVS. Their light curves are  presented  with our targets in Figure  \ref{figlc}  
which shows how the three dC stars are variable in a similar way and with similar amplitude as the  two CH variable giants.
 
 \onlfig{5}  {
   \begin{figure*}
  \centering
 \includegraphics[width=\textwidth  ]{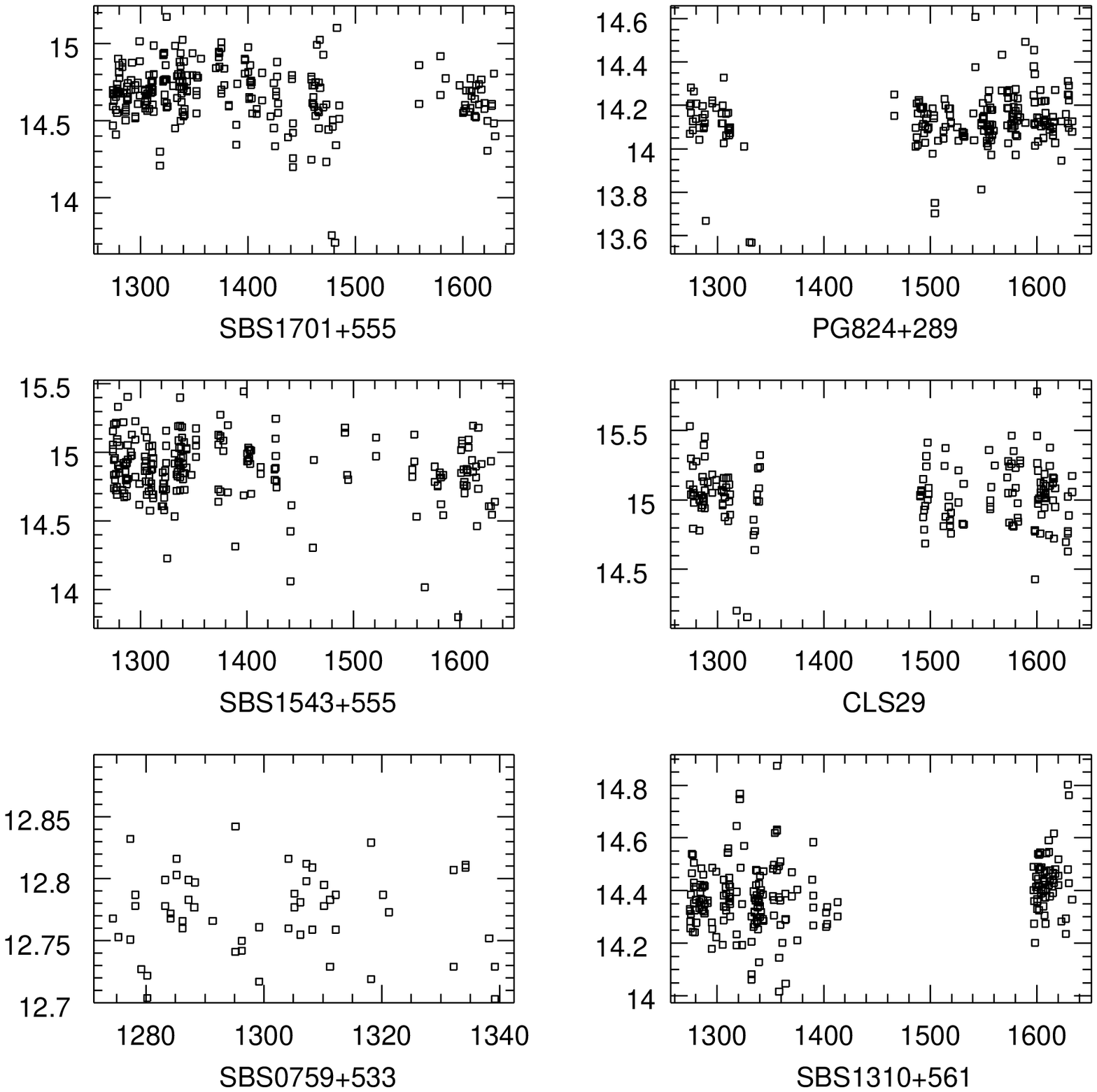}  
      \caption{Light curves   of  our CH stars  plus CLS29 and PG084+289. The dates are MJD-50\,000 days; the data are from the NSVS archive.
	The typical  uncertainties are 0.03 mag for SBS\,0759+533; 0.08 mag for SBS\,1310+561 and PG824+289; 0.11 mag for the other stars. 
              }
         \label{figlc}
   \end{figure*}
 } 
 
\subsection{     Looking for a companion for  SBS\,1310+561 }
It is important to remember that, since according to the evolutionary models no carbon is produced by main sequence hydrogen burning stars, the most reasonable, widely accepted explanation is that dCs belong to evolved binary systems; the presently observed carbon stars are thought to be formed through mass transfer of carbon enriched material while the initial primary component was ascending the asymptotic giant branch as a C giant  (Green \cite{green00} and references therein). 
deKool and Green (\cite{dek95}) constructed several models of binary system evolution leading to dC star formation; their predictions have been confirmed by radial velocity and chemical abundances studies of several targets (Lucatello et al. \cite{luca05}, Behara et al. \cite{beh10}). The characteristics of the individual spectra depend on the distance between the components of the binary system and the nature of the collapsed companion: \\  {
 the prototype G77$-$61 is a single line spectroscopic binary not showing emission lines;  the  companion, invisible in the optical tange  is a WD with Teff $<$ 6000K 
  (Dearborn et al. \cite{dearb86}).
 In SBS\,1517+5017  no emissions are visible but  the hot white dwarf contributes to the continuum and to the absorption  lines already in the blue spectral region.  
 CLS29  only shows H$\alpha$ in emission ( figure 5 of Totten and Irwin \cite{totten98})
 For PG 0824+289 the Balmer emission lines have been attributed to the heating of the dC atmosphere by the visible WD.
Except for the emission lines, SBS\,1310+561 does not show a composite spectrum. Most probably the visible star has an optically  invisible  companion, 
possibly a close WD. 
This can be suggested by the emission of the \ion{Ca}{ii} H and K lines which are normally indicators of strong chromospheric activity. For a single dwarf star  this should not be the case, but in a binary system 
a white dwarf companion in close orbit can spin up the M dwarf via tidal locking and  thus trigger chromospheric activity on the cool star.
In this respect we remember that  mass transfer or a  similar interaction was  suggested as the cause of the luminosity fluctuations for CLS29 and PG0824+289.
  
  Ultraviolet   spectra  might confirm  the presence of a white dwarf companion and hence the binarity of SBS\,1310+561.   We can compare the photometry only with the SLOAN  data for similar stars. 
  The average  $u-g$  computed from  65 stars classified  "D" in table 1  of Downes et al (\cite {down04})  gives  $u-g$=2.7 with  $\sigma$=0.6. 
   For this star  $u-g$ = 2.3. }
We finally  verified that the Galex archive (Martin et al. \cite{martin05})  (farUV$\sim$1400$-$1800, nearUV$\sim$1800$-$2800\,\AA) contains data of our targets, of several known dCs and of the FHLC present  in the list of Totten el al. (\cite{totten00}). Excluding the dwarfs with composite spectra, most of the stars, including SBS\,1543+555, with similar spectral type and  similar $B$ magnitudes  or brighter than SBS\,1310+561 are not  detected. Very few, including SBS\,0759+533 and SBS\,1701+555, have the difference NUV$-B\ \sim6.5$ mag. For the two dCs G77$-$61 and CLS50 the difference  is $\sim 5.3$ mag. SBS\,1310+561 has the smallest difference NUV$-B = 3.7 \pm 0.3$ mag, possibly indicating the presence of a low luminosity but hot source.

At present we do not have enough information for deeper speculations: there are no other data published for any of these stars in the near or in the far IR, where an emission excess could witness the presence of a residual debris disk from the mass transfer event (see Fazio \& Lowrance \cite{fazio08}).

\subsection{  Luminosities and distances }

From the data at our disposal we could estimate approximate values of the distance to our targets. For the dC star SBS\,1310+561 we started assuming for the absolute magnitude in the K band a value  $+6.1 < MK_s < +6.6$ which is the range for dC stars with known parallaxes (Lowrance et al. \cite{low03} and references therein). With the observed $K_s$=+10.80 mag the distance modulus is 
 then $4.2 <$ DM $< 4.7$ and the distance  $70 < $d$ < 87$ pc. 

To compute the absolute visual magnitude we must take into account the variability of the star. We adopted $V = 14.6$ mag, average between our $V = 14.76$ and
 $V = 14.44$ from GSC2.3 and the two limits for the distance modulus obtained from the K magnitude which is less sensitive to variability. These data yield the absolute visual magnitude $M_V = +10.1 \pm 0.2$, in good agreement with the mean value of the dC stars with measured parallaxes. Finally we can give a crude estimate of the space velocity components and try to put some constraints on the Galactic population membership of SBS\,1310+561. For the mean value of the distance d=80 pc, the parallax is $\pi$ = 12.5 mas. By combining this value with the kinematical properties of the star yields the space velocity components (U,V,W) = (-6$\pm$5, -10$\pm$ 8, -30$\pm$15) km/s,  consistent with disk membership (in Galactocentric coordinates : -6, +210, -30 km/s). The values are corrected for the solar motion (10, 15, 7)km/s with respect to the LSR; U is positive in the direction of the Galactic center. The formulae  used are those described by  Johnson \& Soderblom (\cite{jo87}). 

To compute the absolute magnitudes $M_K$  and the distances to the three CH stars we used the empirical fitting formula:
\begin{equation}
\ \ Log(M_{K}  +9.0) = 1.14 - 0.65(J - K)  
\end{equation}
obtained by Totten et al. (\cite{totten00})  from a selected sample of giants
 (standard deviation =0.5mag) and applied to  all their faint high latitude carbon stars. 
The computed   $K$-band absolute magnitudes  in the SAAO system and the derived heliocentric distances in  are
 given in Table \ref{tab4}. 
 
    Assuming the standard deviation of fitting formula  as the main cause of the uncertainty on $M_{K}$ and therefore on the distance modulus, the uncertainties on the distances are about 10\%.
 In the 2MASS system the absolute magnitudes would differ by about 0.04 mag. 


\begin{table}
\caption{Absolute $K$ magnitudes and distances to the three CH giants. } 
\label{tab4}      
\centering          
\begin{tabular}{ c c c  c}     
\hline\hline       
SBS & $M_K$ & d & $ \Delta$ d \\ 
number & mag & kpc & kpc   \\ 
\hline                    
 0759+533  &  -4.9  & 10 & 1.\\     
 1543+555 &   -5.8 & 20 & 2. \\   
 1701+555  &  -4.3 & 20 & 2. \\   
\hline                  
\end{tabular}
\end{table}

\section{  Summary and Conclusions}

We presented moderate resolution CCD spectra and new photometric data for four SBS candidate C stars. Spectra and colours are consistent with early CH type classification. The most important result is the discovery that SBS\,1310+561 belongs to the small group of dwarf carbon stars. In spite of the limited sample we could verify  the increasing evidence that spectroscopy or colours alone are inconclusive as luminosity discriminants for CH$-$type stars. In fact, only the high proper motion, and possibly the strength of the red CaH bands lead to the conclusion that SBS\,1310+561 is a main sequence star, an extremely rare case of dwarf carbon showing Balmer and \ion{Ca}{ii} lines in emission. Dealing with a possible close binary system close time spaced  observations 
 should  be performed to shed more light on the object which shows photometric fluctuations, but the sparse  data   do not allow to speculate too much  about a  possible light curve and/or periodical spectral variations. 

\begin{acknowledgements}
 K.S.G. and C.R. would like to thank the Bologna  and Asiago Observatories staff for the logistic support and the technical assistance during the observations. The University La Sapienza supported the project with funds from the MIUR.
  This research has made use of SIMBAD database operated at CDS, Strasbourg, France, and also of the Two Micron All$ -$Sky Survey database, which is a joint project of the University of Massachusetts and the Infrared Processing and Analysis  Center/California Institute of Technology. 
GALEX is operated for NASA by the California Institute of Technology under NASA contract NAS5-98034. 
IRAF is distributed by the NOAO which is operated by AURA under contract with NFS. 
\end{acknowledgements}

\end{document}